\documentclass[a4paper,twocolumn]{article}

\usepackage[utf8]{inputenc}
\usepackage[T1]{fontenc}
\usepackage{float}
\usepackage{amsmath,amssymb}
\usepackage{mathrsfs}
\usepackage{theorem}
\usepackage{graphicx}
\usepackage{color}
\usepackage{wasysym}
\usepackage{hyperref} 
\usepackage{verbatim}

\definecolor{blue}{rgb}{0,0,1}
\definecolor{green}{rgb}{0,0.65,0.5}
\definecolor{verde}{rgb}{0.,.5,0.4}
\definecolor{marron}{rgb}{0.7,0.2,0.1}
\definecolor{red}{rgb}{1,0,0}
\definecolor{vio}{rgb}{1,0,1}
\definecolor{ama}{rgb}{1,1,0}

\newcommand{\bc}{\begin{center}}
\newcommand{\ec}{\end{center}}
\newcommand{\be}{\nopagebreak[3]\begin{equation}}
\newcommand{\ee}{\end{equation}}
\newcommand{\ba}{\nopagebreak[3]\begin{eqnarray}}
\newcommand{\ea}{\end{eqnarray}}

{\theorembodyfont{\sffamily} }
{\theorembodyfont{\sffamily} }
{\theorembodyfont{\sffamily} }

\begin{document}

\title{\bf 
Single time dynamical model for \\
equations of motion 
of relativistic retarded systems
}

\author{
Osvaldo M. Moreschi\thanks{Email: o.moreschi@unc.edu.ar}
\\
{\rm \small Facultad de Matemática Astronomía, Física y Computación (FaMAF),} \\
{\rm \small Universidad Nacional de C\'{o}rdoba,}\\
{\rm \small Instituto de F\'\i{}sica Enrique Gaviola (IFEG), CONICET,}\\
{\rm \small Ciudad Universitaria, (5000) C\'{o}rdoba, Argentina.}
}

\maketitle

\begin{abstract}

We present a procedure to build a single time model
for the equations of motion of relativistic retarded systems
composed of several particles; at any desired level of accuracy.
We treat the especial case of a binary system.

We apply this model to the classical electromagnetic binary
particle system.

We mentioned some differences with previous approaches
and discussed the implications for linear 
gravitational models.

\end{abstract}


\vspace{5mm}


\section{Introduction}

\subsection{Content of this article}

Models for retarded particle systems that use a single dynamical time
are very convenient, since the numerical calculation simplifies
considerably; instead to deal with several proper times and
exact retarded effects.
The way in which one approaches this procedure influences the
precision and predictive power of the final dynamical equations.
It has been customary to emphasize either an expansion in terms
of interaction constants and/or in terms of the velocity of
the particles.
We here instead emphasize the precision in the calculation 
of the retarded quantities involved; which provides a new
perspective into this topic.

The subject of this work is of interest to the dynamics of charged
particles in Minkowski spacetime and to models for the motion of black holes,
when treated as `particles' in first order of the field equations. 
In this setting, then each particle feels the retarded fields generated 
by the other particle, with respect to a flat background metric.
Although we will have in mind a binary system, the study can
be extended to any number of particles.

The problem treated here has been studied in the past by several authors;
but normally severe approximations have been made that
had as a consequence that the retarded effects had a crude
estimation. 
We instead, see the problem from a different perspective;
so that for each dynamical model for a physical system,
we take the forces as the source of information;
and then we try to build an approximation
that calculates the retarded effects with the desired accuracy.
For this reason we will obtain equations of motions that
differ from previous results.

When constructing a single time dynamical model for
equations of motion of relativistic retarded systems one
would like to reproduce as precisely as possible the
global properties of the original retarded system.
In our approach the aim is to avoid the introduction of cumulative
effects on the relativistic dynamics, due to inaccurate retarded times,
positions and velocities 
calculations. These calculations are independent of the nature of
the theory, the field equations, or of the forces; they just take into
account the Lorentzian relativistic nature of past null cones.

Since we are showing a new point of view to an old problem,
we are going through a detailed presentation of the involved topics.
We first, in section \ref{sec:retarded-binary}, concentrate on the calculations
of the retarded times in Minkowski spacetime; and use them to approximate the retarded
forces by a set that depends on a single dynamical time.
We present a procedure that only takes into account the value of
the positions and velocities of the particles and also use
a single evaluation of the forces; which we call the order one calculation.

In section \ref{sec:lagrangian} we discuss the possibility to approximate
the dynamics by a Lagrangian system.

The order two retarded forces are calculated in section \ref{sec:order2}\,;
where we apply a Runge-Kutta like approach.

We present the forces for a charged binary system
in section \ref{sec:electro}\,.

The final section is reserved for comments of our approach
and its relation with previous works.

Although we have in mind a binary system, whose components we label with $A$ and $B$;
in order to simplify the reading, whenever possible, when theres is no room for ambiguities,
we neglect the corresponding subindices $A$ and $B$.


\section{Retarded effects in a binary system}\label{sec:retarded-binary}

\subsection{The retarded times issue}

We assume the dynamics of a relativistic binary system is determined from 
the equations of motion of the form:
\begin{equation}\label{eq:motA}
m_A \frac{d \vec v_A}{dt} = \vec f_A(\vec r_A(t), \vec v_A(t)
; \vec r_B(t_\text{rB}) ,\vec v_B(t_\text{rB}) ,\vec a_B(t_\text{rB}))
,
\end{equation}
and
\begin{equation}\label{eq:motB}
m_B \frac{d \vec v_B}{dt} = \vec f_B(\vec r_B(t), \vec v_B(t)
; \vec r_A(t_\text{rA}) ,\vec v_A(t_\text{rA}) ,\vec a_A(t_\text{rA}) )
;
\end{equation}
where $t_\text{rA}$ and $t_\text{rB}$ are the corresponding retarded times.

Then, the main objective is to obtain a new approximate
version of (\ref{eq:motA}) and (\ref{eq:motB})
which will only include reference to the coordinate time $t$;
and any appearance of quantities evaluated at retarded times has been replaced
by an appropriate expansion, 
in terms of a number of evaluations of the forces used to
calculate the retarded times.

One can see that
to determine the force on body $A$ one has to know the retarded position,
velocity and acceleration
of body $B$ at $t_\text{rB}$; but we have the information $\vec r_B(t)$,
$\vec v_B(t)$. In turn, to calculate the time derivative of the velocity
for body $B$ one needs to know the retarded position, velocity and acceleration 
of body $A$;
but we have the information $\vec r_A(t)$,
$\vec v_A(t)$.

Using a four dimensional notation in which $z_B^\mu(\tau_B)$ denotes
the position of particle $B$, in terms of a Cartesian coordinate system,
 at proper time $\tau_B$, and $y_A^\mu$
the field position of particle $A$,
one notes that since the point $z_B^\mu(\tau_B)$ is in the past null cone of $y_A^\mu$,
one has
\begin{equation}
 y_A^0  - z_B^0(\tau_B) = | y_A^i  - z_B^i(\tau_B) | ;
\end{equation}
where $||$ means the modulus of the spacelike relative position vector in
the Cartesian frame one is using; that we could call $\hat r(\tau_B)$.

We would like to refer to the binary system in terms of a common coordinate
time, so that we will use:
\begin{equation}
t - t_\text{rB} = |\vec r_A(t) - \vec r_B(t_\text{rB})|/c ,
\end{equation}
and
\begin{equation}
t - t_\text{rA} = |\vec r_B(t) - \vec r_A(t_\text{rA})|/c .
\end{equation}
These can also be expressed in terms of $\Delta t_B =t - t_\text{rB}$
and $\Delta t_A =t - t_\text{rA}$ as
\begin{equation}\label{eq:deltatb}
\Delta t_B = |\vec r_A(t) - \vec r_B(t - \Delta t_B)|/c ,
\end{equation}
and
\begin{equation}\label{eq:deltata}
\Delta t_A = |\vec r_B(t) - \vec r_A(t- \Delta t_A)|/c .
\end{equation}

Therefore, in order to be able to calculate accurately the retarded times,
we need to know the trajectories to the past of the coordinate time $t$
with the required precision.
Different approaches to do this are mentioned below.

In reference \cite{Darwin1920} the author estimates
the retarded times by approximating equations (\ref{eq:deltatb}) and (\ref{eq:deltata})
by quadratic equations; where it was assumed an expansion in terms of
forces, or accelerations.
The approach used in the Landau-Lifshitz textbook\cite{Landau75} is based on the idea
that if the coordinate velocities of system $B$ are small, then its configuration will
not change significantly during the time $\hat r(\tau_B)/c$ (where we introduce explicitly
the velocity of light $c$). Then it is natural to think in the coordinate retarded time
$z_B^0$ as
\begin{equation}
z_B^0 = y_A^0  - \hat r/c ;
\end{equation}
and expand any reference to $z_B^0$ in series of powers of $\hat r/c$.
Let us call $t=y_A^0$ de value of the time coordinate of one particle, 
then the fields of the other particle are evaluated at the retarded time
$t_\text{r} = z_B^0 = t - \hat r(t,t_\text{r})/c$.
Due to the small velocity assumption one can expand any field as
\begin{equation}
\begin{split}
\mathcal{F}(t_\text{r}, \vec r_B(t_\text{r}) )
=&
\mathcal{F}(t, \vec r_B(t) )
- \frac{\hat r}{c}  \frac{\partial \mathcal{F}}{\partial t}  \\
&
- \frac{\hat r}{c} \nabla_{\vec r_B(t)} \mathcal{F} \cdot \vec v_B  
+ \mathscr{O}(\frac{\hat r}{c} )^2
;
\end{split}
\end{equation}
which is in fact a Taylor expansion around the fields evaluated at time $t$
to the time $t' = t -\Delta t$
with $\Delta t \approx  \frac{\hat r}{c}$; but of course, $\frac{\hat r}{c}$
need not be small in any sense. It is only assumed that the velocities are
small.
They apply this to a smooth energy-momentum tensor; but the procedure
does not seem to have a regular behavior when one takes the limit
to a Dirac delta distribution.
Our approach differs from these works in that our primary objective
is the precision for the calculation of the retarded quantities;
which is gained at the expenses of evaluations of the force terms.
This is in contrast to fix first an order for the evaluations of the
forces and then calculate the retarded effects.
In any case, since we are concerned with numerical evaluation of the
dynamics, we next organize the presentation in terms of the
number of evaluations of the forces.

\subsection{Order one retarded approach}
We would like to think of this problem by concentrating
in a point like bodies approach.
The guiding idea is to obtain an approximate expression,
that uses all the available kinematical and dynamical information at
 a common time coordinate $t$,
instead of the proper times of each particles.

By order one retarded effects we mean those that can be calculated
using one evaluation of the forces for each trajectory,
using the kinematical data at our disposal at time $t$;
namely positions and velocities of the particles.
More specifically we approximate the trajectory into the past by the 
corresponding parabolic motion determined by the acceleration at the
present time $t$.

Then, in turn, 
by order two retarded calculation of the retarded time
we mean those that are calculated by
making two evaluations of the forces in each trajectory.
However it must be noted that when using this information to 
evaluate of the force of one particle, it will include the 
evaluation of the force of the other particle;
in order to estimate the retarded effects.
Then, to avoid the inclusion of separate order keeping, we
will also say that the order of the force calculated
with an order $n$ retarded time, will also be called of order $n$.

Let us think in an iterative procedure such that we start by evaluating
a zero order retardation effect force by assuming a linear
trajectory to the past; namely:
Let us call $t_\text{rB0} = t - \Delta t_{B0}$ the solution of
\begin{equation}\label{eq:deltat-b1}
\begin{split}
\Delta t_{B0} =&
t - t_\text{rB0} \\
=& |\vec r_A(t) - \big(\vec r_B(t) 
-(t - t_\text{rB0}) \vec v_B(t) \big)|/c \\
=& |\vec r_A(t) - \big(\vec r_B(t) 
- \Delta t_{B0} \vec v_B(t) \big)|/c \\
=& 
\bigg(
r_{AB}^2 
- 2 \Delta t_{B0} \big(\vec r_A(t) - \vec r_B(t) \big)	\cdot \vec v_B(t) \\
&+ \Delta t_{B0}^2 v_B(t)^2
	\bigg)^{1/2}/c 
;
\end{split}
\end{equation}
which is the solution to the intersection of a linear motion with
the past null cone of $\big(t,\vec r_A(t)\big)$.
We also define $t_\text{rA0} = t - \Delta t_{A0}$ to be the solution of
\begin{equation}\label{eq:deltat-a1}
\begin{split}
\Delta t_{A0} =&
t - t_\text{rA0} \\
=& |\vec r_B(t) - \big(\vec r_A(t) 
-(t - t_\text{rA0}) \vec v_A(t) \big)|/c \\
=& |\vec r_B(t) - \big(\vec r_A(t) 
- \Delta t_{A0} \vec v_A(t) \big)|/c 
.
\end{split}
\end{equation}
Expressing \ref{eq:deltat-b1} as a quadratic equation, namely
\begin{equation}\label{eq:deltat-b1q}
\begin{split}
\Delta t_{B0}^2 \big(1  - v_B(t)^2 / c^2\big)
+& 2 \Delta t_{B0} \big(\vec r_A(t) - \vec r_B(t) \big)	\cdot \vec v_B(t)/c^2 \\
-& r_{AB}^2 / c^2
= 0
;
\end{split}
\end{equation}
one has the solution
\begin{equation}\label{eq:deltat-b1-b}
\begin{split}
\Delta t_{B0} = &
\frac{
\sqrt{\big(1  - \frac{v_B^2}{c^2} \big) \frac{r_{AB}^2}{c^2} + \frac{(r_{AB}v_B)^2}{c^4}   }
-	\frac{(r_{AB}v_B)}{c^2} 
	}{\big(1  - \frac{v_B^2}{c^2}\big)}
;
\end{split}
\end{equation}
and similarly for $\Delta t_{A0}$.
The notation we are using is:
$\vec r_{AB}(t) = \vec r_A(t) - \vec r_B(t)$,
$r_{AB} = |\vec r_{AB}(t)|$,
for the scalar product
$(r_{AB}v_B) = \vec r_{AB}(t) \cdot \vec v_B(t)$
and
$v_B = |\vec v_B(t)|$.

Then, we define the zero order retarded forces as
\begin{equation}\label{eq:fa1}
\begin{split}
\vec {f}_{A0}(\vec r_A, \vec v_A ; t ) = \vec {f}_A\Big(& \vec r_A , \vec v_A 
; \vec r_B(t) - \Delta t_{B0}  \vec v_B(t) , 
\vec v_B(t) , 0
\Big)
,
\end{split}
\end{equation}
and
\begin{equation}\label{eq:fb1}
\begin{split}
\vec {f}_{B0}(\vec r_B, \vec v_B ;t ) = \vec {f}_B\Big(& \vec r_B, \vec v_B 
; \vec r_A(t) - \Delta t_{A0}  \vec v_A(t), 
\vec v_A(t) ,  0
\Big)
;
\end{split}
\end{equation}
that is, in this first stage we assume the linear motion.
With this we define $t_\text{rB1} = t - \Delta t_{B1}$ the solution of
\begin{equation}
\begin{split}
t - t_\text{rB1} =& |\vec r_A(t) - \big(\vec r_B(t) 
-(t - t_\text{rB1}) \vec v_B(t) \big) \\
&+ \frac{(t - t_\text{rB1})^2}{2 m_B} \vec {f}_{B0}(t, \vec r_B, \vec v_B)
|/c ;
\end{split}
\end{equation}
or
\begin{equation}
\begin{split}
\Delta t_{B1} =& |\vec r_A(t) - \big(\vec r_B(t) 
- \Delta t_{B1} \vec v_B(t) \big) \\
&+ \frac{\Delta t_{B1}^2}{2 m_B} \vec {f}_{B0}(t, \vec r_B, \vec v_B)
|/c ;
\end{split}
\end{equation}
which is the solution to the intersection of a quadratic motion with
the past null cone of $\big(t,\vec r_A(t)\big)$.
If one expresses $\Delta t_{B1} = \Delta t_{B0} + \delta t_{B1}$, one can see that
$\delta t_{B1} = \mathscr{O}(\vec {f}_{B0}/m_B)$; so that in first order
of $\mathscr{O}(\vec {f}_{B0}/m_B)$ one could instead solve for
\begin{equation}\label{eq:tB11}
\begin{split}
\Delta t_{B1(1)} =& |\vec r_A(t) - \big(\vec r_B(t) 
- \Delta t_{B1(1)} \vec v_B(t) \big) \\
&+ \frac{\Delta t_{B0}^2}{2 m_B} \vec {f}_{B0}(t, \vec r_B, \vec v_B)
|/c ;
\end{split}
\end{equation}
since $\Delta t_{B1(1)}-\Delta t_{B1} = \mathscr{O}(\vec {f}_{B0}^2/m_B^2)$.
In this way one again can deal with a quadratic equation which is simpler to handle.
But conceptually we would like to deal with the exact solution for this
retarded time, and since all this calculation will end up to be carried
out numerically, we could resort to an iterative scheme of the form
\begin{equation}
\begin{split}
\Delta t_{B1(n+1)} =& |\vec r_A(t) - \big(\vec r_B(t) 
- \Delta t_{B1(n+1)} \vec v_B(t) \big) \\
&+ \frac{\Delta t_{B1(n)}^2}{2 m_B} \vec {f}_{B0}(t, \vec r_B, \vec v_B)
|/c ;
\end{split}
\end{equation}
for $n=1,2,3,...$; whose solutions are expresses in the form
of \ref{eq:deltat-b1-b}, namely
\begin{equation}\label{eq:deltat-b2}
\begin{split}
\Delta t_{B1(n+1)} = &
\frac{
	\sqrt{\big(1  - \frac{v_B^2}{c^2} \big) \frac{r_{AfnB}^2}{c^2} + \frac{(r_{AfnB}v_B)^2}{c^4}   }
	-	\frac{(r_{AfnB}v_B)}{c^2} 
}{\big(1  - \frac{v_B^2}{c^2}\big)}
;
\end{split}
\end{equation}
where $\vec r_{AfnB} = \vec r_A(t) + \frac{\Delta t_{B1(n)}^2}{2 m_B} \vec {f}_{B0}(t, \vec r_B, \vec v_B)-\vec r_B(t)$.
Of course one could also use the analytic expressions for the solutions of
quartic polynomial equations. 
It is clear that
the previous iterative scheme will produce the desired solution with a simplification
of a numerical code.

Similarly we define $t_\text{rA1} = t - \Delta t_{A1}$ to be the solution of
\begin{equation}
\begin{split}
t - t_\text{rA1} =& |\vec r_B(t) - \big(\vec r_A(t) 
-(t - t_\text{rA1}) \vec v_A(t) \big) \\
&+ \frac{(t - t_\text{rA1})^2}{2 m_A} \vec {f}_{A0}(t, \vec r_A, \vec v_A)
|/c .
\end{split}
\end{equation}

Then, we define the first order retarded forces by
\begin{equation}\label{eq:fAsecond}
\begin{split}
\vec {f}_{A1}(\vec r_A, \vec v_A ; t ) =& \vec {f}_A(\vec r_A , \vec v_A 
; \\
& \vec r_B(t) - \Delta t_{B1} \, \vec v_B(t) + \frac{\Delta t_{B1}^2}{2 m_B} \vec {f}_{B0}, \\
& \vec v_B(t) - \frac{\Delta t_{B1}}{m_B}  \, \vec {f}_{B0} , \\
& \frac{1}{m_B} \vec {f}_{B0}
)
,
\end{split}
\end{equation}
and
\begin{equation}\label{eq:fBsecond}
\begin{split}
\vec {f}_{B1}(\vec r_B, \vec v_B ; t ) =& \vec {f}_B(\vec r_B, \vec v_B 
; \\ 
& \vec r_A(t) - \Delta t_{A1} \, \vec v_A(t) + \frac{\Delta t_{A1}^2}{2 m_A} \vec {f}_{A0}, \\
& \vec v_A(t) - \frac{\Delta t_{A1}}{m_A} \, \vec {f}_{A0} \\
& \frac{1}{m_A} \vec {f}_{A0}
)
.
\end{split}
\end{equation}

This is an improvement to the calculations carried out in
\cite{Darwin1920}, since in that reference he neglects higher order
effects of the forces.
That is, he also applies a Taylor expansion of the fields
in the Lagrangian, around the time $t$ 
to the retarded time $t' = t -\Delta t$, for each particle;
but $\Delta t$ is calculated with less accuracy and
the accelerations are calculated at time $t$.

Suppose that one would like to calculate the dynamic evolution
numerically.
We have seen in this procedure that we are taking as initial
data the positions and velocities; which at this stage are
taken as exact quantities. With these we estimate two set of
quantities: the retarded times and the forces taking into account
retarded effects.
When integrating the corresponding equations of motions one
will deal with errors in the trajectories, that is in position
and velocities, which where non existent in the initial data.
This indicates the important role played by the errors
in the determination of retarded times and first order
retarded forces; since the numerical calculation should
have a precision according to the quality of the
calculation of the retarded time and forces.
We will return to this issue in section \ref{sec:order2}.

\paragraph{Summary:}
	
We can express this approach in a way 
that facilitates the writing of a numerical algorithm.
Let $\vec r_A$, $\vec v_A$, $\vec r_B$ and $\vec v_B$ be the
position and velocity vectors in the
three dimensional Cartesian system at time $t$.
Then let us define the zero order retarded lapse of times
\begin{equation}\tag{\ref{eq:deltat-b1-b}}
\begin{split}
\Delta t_{B0} = 
\frac{
	\sqrt{\big(1  - \frac{v_B^2}{c^2} \big) \frac{r_{AB}^2}{c^2} + \frac{(r_{AB}v_B)^2}{c^4}   }
	-	\frac{(r_{AB}v_B)}{c^2} 
}{\big(1  - \frac{v_B^2}{c^2}\big)}
,
\end{split}
\end{equation}
\begin{equation}\label{eq:deltat-a1-b}
\begin{split}
\Delta t_{A0} = 
\frac{
	\sqrt{\big(1  - \frac{v_A^2}{c^2} \big) \frac{r_{AB}^2}{c^2} + \frac{(r_{AB}v_A)^2}{c^4}   }
	+	\frac{(r_{AB}v_A)}{c^2} 
}{\big(1  - \frac{v_A^2}{c^2}\big)}
.
\end{split}
\end{equation}
The zero order retarded forces are defined by
\begin{equation}\tag{\ref{eq:fa1}}
\begin{split}
\vec {f}_{A0}(\vec r_A, \vec v_A  ; t) = \vec {f}_A\Big(& \vec r_A , \vec v_A 
; \vec r_B(t) - \Delta t_{B0}  \vec v_B(t) , 
\vec v_B(t) , 0
\Big)
,
\end{split}
\end{equation}
and
\begin{equation}\tag{\ref{eq:fb1}}
\begin{split}
\vec {f}_{B0}(\vec r_B, \vec v_B ; t ) = \vec {f}_B\Big(& \vec r_B, \vec v_B 
; \vec r_A(t) - \Delta t_{A0}  \vec v_A(t), 
\vec v_A(t) ,  0
\Big)
.
\end{split}
\end{equation}
Then, we define $\Delta t_{B1} = t - t_\text{rB1}$ to be the 
appropriate iterative solution of
\begin{equation}\tag{\ref{eq:deltat-b2}}
\begin{split}
\Delta t_{B1(n+1)} = &
\frac{
	\sqrt{\big(1  - \frac{v_B^2}{c^2} \big) \frac{r_{AfnB}^2}{c^2} + \frac{(r_{AfnB}v_B)^2}{c^4}   }
	-	\frac{(r_{AfnB}v_B)}{c^2} 
}{\big(1  - \frac{v_B^2}{c^2}\big)}
;
\end{split}
\end{equation}
for $n=0,1,2,3,...$, 
where 
\begin{equation}
\vec r_{AfnB} = \vec r_A(t) + \frac{\Delta t_{B1(n)}^2}{2 m_B} \vec {f}_{B0}(t, \vec r_B, \vec v_B)-\vec r_B(t)
;
\end{equation}
with $\Delta t_{B1(0)} =\Delta t_{B0}$.
Similarly, we define $\Delta t_{A1} = t - t_\text{rA1}$ to be the 
appropriate iterative solution of
\begin{equation}\label{eq:deltat-a2}
\begin{split}
\Delta t_{A1(n+1)} = &
\frac{
	\sqrt{\big(1  - \frac{v_A^2}{c^2} \big) \frac{r_{BfnA}^2}{c^2} + \frac{(r_{BfnA}v_A)^2}{c^4}   }
	-	\frac{(r_{BfnA}v_A)}{c^2} 
}{\big(1  - \frac{v_A^2}{c^2}\big)}
;
\end{split}
\end{equation}
where 
\begin{equation}
\vec r_{BfnA} = \vec r_B(t) + \frac{\Delta t_{A1(n)}^2}{2 m_A} \vec {f}_{A0}(t, \vec r_A, \vec v_A)-\vec r_A(t)
;
\end{equation}
with $\Delta t_{A1(0)} = \Delta t_{A0}$.
Then, the order one forces single time model of the retarded system 
(\ref{eq:motA}),  (\ref{eq:motB}) can be expressed as:
\begin{equation}\tag{\ref{eq:fAsecond}}
\begin{split}
\vec {f}_{A1}(\vec r_A, \vec v_A ; t ) =& \vec {f}_A(\vec r_A , \vec v_A 
; \\
& \vec r_B(t) - \Delta t_{B1} \, \vec v_B(t) + \frac{\Delta t_{B1}^2}{2 m_B} \vec {f}_{B0}, \\
& \vec v_B(t) - \frac{\Delta t_{B1}}{m_B}  \, \vec {f}_{B0} , \\
& \frac{1}{m_B} \vec {f}_{B0}
)
,
\end{split}
\end{equation}
and
\begin{equation}\tag{\ref{eq:fBsecond}}
\begin{split}
\vec {f}_{B1}(\vec r_B, \vec v_B ; t ) =& \vec {f}_B(\vec r_B, \vec v_B 
; \\ 
& \vec r_A(t) - \Delta t_{A1} \, \vec v_A(t) + \frac{\Delta t_{A1}^2}{2 m_A} \vec {f}_{A0}, \\
& \vec v_A(t) - \frac{\Delta t_{A1}}{m_A}  \, \vec {f}_{A0} \\
& \frac{1}{m_A} \vec {f}_{A0}
)
.
\end{split}
\end{equation}
	
It should be emphasized that in order to minimally improve in the
calculation of the forces, using the universal time $t$,
and taking into account first order retarded effects,
one must make an evaluation of the force in the argument of
the corrected positions and velocities. This is completely missing in the 
Darwin\cite{Darwin1920}
and Landau-Lifshitz\cite{Landau75} approaches.
In fact, it is not at all clear at this stage that there exists a normal
Lagrangian formulation of this system.

\section{Possibility of a Lagrangian treatment of the retarded effects}\label{sec:lagrangian}

Let us discuss the situation in which, if system $B$ where given,
then there exists a Lagrangian $L_A$ for particle $A$
of the form
\begin{equation}\label{eq:lag_L_A}
L_A = L_A(\vec r_a(t) , \vec v_A(t), \vec r_B(t-\Delta t_B) , \vec v_B(t-\Delta t_B) )
;
\end{equation}
where $t_{\text{r}B} = t-\Delta t_B$ is the retarded time of particle $B$
as seen from particle $A$ at $\vec r_A(t)$, as discussed above.

Then, using the procedure described in the previous section, we will
obtain appproximate expressions
\begin{equation}\label{eq:r_B}
\begin{split}
\vec r_B(t-\Delta t_B) = &
\vec r_B(t) - \Delta t_{B} \vec v_B(t) \\
&+ \frac{\Delta t_{B}^2}{2 m_B} \vec {f}_{B}(\vec r_B, \vec v_B; t) 
+ \mathscr{O}(\text{r}2 )
,
\end{split}
\end{equation}
and
\begin{equation}\label{eq:v_B}
\vec v_B(t-\Delta t_B) = \vec v_B(t) -   \frac{\Delta t_{B}}{m_B} \vec {f}_{B}(\vec r_B, \vec v_B; t)
+ \mathscr{O}(\text{r}2 )
;
\end{equation}
where $\mathscr{O}(\text{r}2 )$ means order two in retarded effects
and where $\Delta t_{B}$ and $\vec {f}_{B}(\vec r_B, \vec v_B; t)$ are normally given in terms of 
approximations.

Note that we have used explicitly the appearance of force terms,
although the starting equations are $\frac{d \vec v_B(t)}{dt} = \frac{1}{m_B} \vec {f}_{B}$.
Therefore, in looking for a Lagrangian that approximates the system
(\ref{eq:motA}) and (\ref{eq:motB}) we observe two attitudes:
one in which one uses in the above expressions $\frac{d \vec v_B(t)}{dt}$,
and the other in which one uses $\frac{1}{m_B} \vec {f}_{B}(\vec r_B, \vec v_B; t)$.

When using the first approach, one is concerned 
with the apparition, in the Lagrangian, of terms of the form
\begin{equation}
\frac{d \vec v_B}{dt} \cdot \vec r_A ,
\end{equation}
and
\begin{equation}
\frac{d \vec v_B}{dt} \cdot \vec v_A .
\end{equation}
The first type might probably be dealt with by adding to the Lagrangian
a new term involving the time derivative of $\vec v_B \cdot \vec r_A$.
But the second type, seems much more difficult to handle.
Terms of this type seem to pose an obstruction for building the
desired Lagrangian with this approach.

Instead when using the second approach, the introduction of expressions
involving $\frac{1}{m_B} \vec {f}_{B}(\vec r_B, \vec v_B; t)$ will not introduce
any problem form the Lagrangian program point of view, since all the expressions
still depend on positions and velocities only. In a sense one is introducing
more precision at the cost of introducing higher order expressions in the
forces; since these expressions appear in the interactions terms of
the Lagrangian.

The first approach was used in reference \cite{Darwin1920}\,; and we have
seen that this Lagrangian model simplifies severely the dynamics,
in several ways.
So, we recommend the second approach based on the approximation
given by (\ref{eq:lag_L_A}), (\ref{eq:r_B}) and (\ref{eq:v_B}),
or their higher order versions;
since, although it introduces higher order force terms in the Lagrangian,
it is based on a more precise approximation to the original
dynamics of equations (\ref{eq:motA}) and (\ref{eq:motB}).

\section{Building an order two retarded approach}\label{sec:order2}

Although we have improved over reference \cite{Darwin1920} by
taking into account the higher order retardation effects on the accelerations,
we still share the shortcoming associated to the fact that
we are using a second order Taylor expansion for the position around time $t$.
It is not clear at this stage what cumulative effects this
will produce in the dynamics of the solutions to the approximate
equations of motion, based on this Taylor expansion.
The question is: suppose that instead to calculate the retarded times
based on a second order Taylor expansions of the positions,
we use an approximation of the trajectory
based on a higher \emph{order of accuracy} Runge-Kutta (R-K) time step calculation; then,
how does the dynamic changes in the equation of motion
based in this new approximation?
Is it possible to improve on the previous estimate, by considering
more evaluations of the force? 
A Runge-Kutta  like method are techniques developed for `first order'
ordinary differential equations.
This in principle would introduce much complication in the calculation
since one has to consider in the analysis the retarded effects
at each step.
But, since the problem we are concerned with is actually a
`second order' differential equation, it is worthwhile
to review the R-K logic, to see if one can improve
in the calculation of the retarded times $\Delta t$,
and in the evaluation of the forces depending on a single
dynamical time.

The R-K  integration methods deal with 
`first order' ordinary differential equations; namely
\begin{equation}
\frac{d \mathbf{x}}{dt} = \mathbf{d}(t, \mathbf{x} )
;
\end{equation}
where $\mathbf{x}$ is defined in terms of position and velocities of the point
like objects\cite{Ralston1978}.

Let us review here the \emph{second order R-K method}\footnote{We will refer 
	to the ``order'' of the R-K method in italics to differentiate it
	from the order of the retarded effects; and we refer to
	`first order' or `second order' of the ordinary differential
	equation between quotes, also to differentiate from the other uses of
	the word ``order''.
	}
 given by:
\begin{equation}\label{eq:rk2}
\mathbf{x}(t+h) = \mathbf{x}(t) + w_1 k_1 + w_2 k_2
;
\end{equation}
with
\begin{eqnarray}
k_1 &=& h \mathbf{d}(t, \mathbf{x}) , \\
k_2 &=& h \mathbf{d}(t+ \alpha h, \mathbf{x}+\beta k_1) 
;
\end{eqnarray}
where we could take\cite{Ralston1978,Cheney:2007:NMC:1537097}
\begin{eqnarray}
\alpha &=& \beta = \frac{2}{3} , \\
w_1 &=& \frac{1}{4} , \\
w_2 &=& \frac{3}{4} .
\end{eqnarray}

In the case of a `second order' ordinary differential equation one has to solve
\begin{equation}
\frac{d^2 \mathbf{y}}{d t^2} = \mathbf{f}(t, \mathbf{y}, \frac{d}{dt}\mathbf{y} )
;
\end{equation}
so that normally one transforms to the `first order' version by defining
\begin{equation}
\mathbf{x} =
\left(
\begin{array}{c}
\mathbf{y} \\ 
\frac{d \mathbf{y}}{dt}
\end{array}
\right) 
;
\end{equation}
so that
\begin{equation}
\mathbf{d}(t, \mathbf{x} ) =
\left(
\begin{array}{c}
\frac{d \mathbf{y}}{dt} \\ 
\mathbf{f}(t, \mathbf{y}, \frac{d}{dt}\mathbf{y} )
\end{array}
\right) 
.
\end{equation}
Then, using the R-K procedure one has
\begin{equation}
\begin{split}
&\left(
\begin{array}{c}
\mathbf{y}(t+h) \\ 
\frac{d \mathbf{y}}{dt}(t+h)
\end{array}
\right) 
=
\left(
\begin{array}{c}
\mathbf{y}(t) \\ 
\frac{d \mathbf{y}}{dt}(t)
\end{array}
\right) 
+
w_1 h \left(
\begin{array}{c}
\frac{d \mathbf{y}}{dt} \\ 
\mathbf{f}(t, \mathbf{y}, \frac{d \mathbf{y}}{dt} )
\end{array}
\right) \\
&+
w_2 h
\left(
\begin{array}{c}
\frac{d \mathbf{y}}{dt} + \beta  h  \mathbf{f}(t, \mathbf{y}, \frac{d \mathbf{y}}{dt} ) \\ 
\mathbf{f}(t+\alpha h, \mathbf{y}+\beta h \frac{d \mathbf{y}}{dt} 
, \frac{d}{dt}\mathbf{y} +\beta h \mathbf{f}(t, \mathbf{y}, \frac{d \mathbf{y}}{dt} ) )
\end{array}
\right) 
;
\end{split}
\end{equation}
that is, for the position one has
\begin{equation}
\begin{split}
\mathbf{y}(t+h) 
& =
\mathbf{y}(t) 
+
w_1 h \frac{d \mathbf{y}}{dt} 
+
w_2 h \big(
\frac{d \mathbf{y}}{dt}(t) 
+  \beta  h  \mathbf{f}(t, \mathbf{y}, \frac{d \mathbf{y}}{dt} )
\big)
;
\end{split}
\end{equation}
from which, in order to agree with the \emph{second order Taylor expansion}, we need
\begin{equation}
w_1 + w_2 = 1 ,
\end{equation}
and
\begin{equation}
w_2 \beta = \frac{1}{2} ;
\end{equation}
which are part of the R-K conditions, but without
requirements on $\alpha$.

In other words, if we had carried out the calculation of the retarded times
$\Delta t$ in terms of a \emph{second order R-K calculation}, we would
have arrived at the same results as above.

But the \emph{order two R-K} scheme also introduces an improvement
on the calculation of the retarded velocity, which enters
as argument in the force at time $t$.
However the new evaluation of the force at time $t+\alpha h = t -\frac{2}{3} \Delta t$,
forces another evaluation of the force of the other particle,
at an earlier retarded time. We could use here the order
one forces, for this task.

\begin{figure}[h]
\includegraphics[clip]{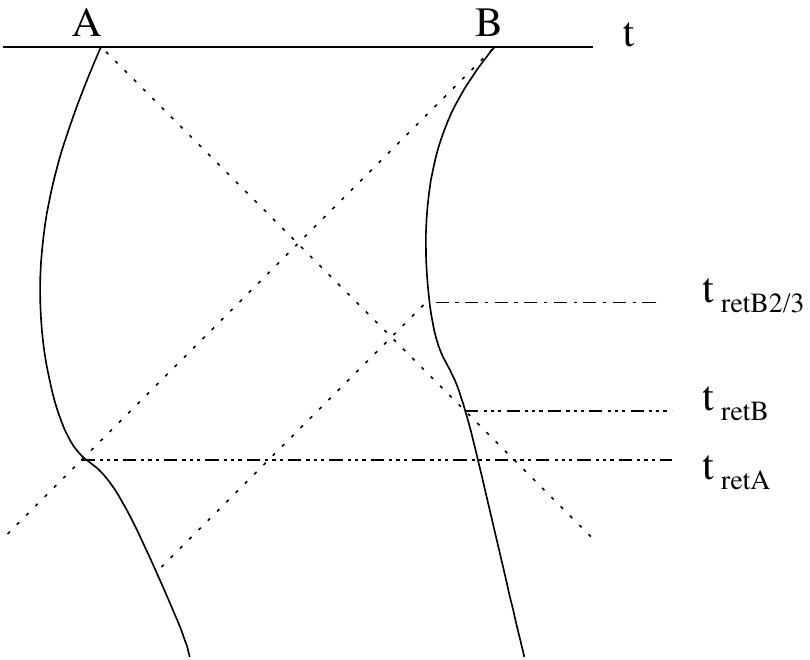}
\caption{Sketch of two arbitrary world lines and corresponding
retarded times. The evaluation of the force, 
for particle $B$, at retarded time $ t -\frac{2}{3} \Delta t$
requires the estimation of the position of particle $A$ at a previous retarded
time $t - \Delta t'_A$, not drawn in the figure.
}\label{fig:ret}
\end{figure}

That is, we now evaluate
\begin{equation}
\begin{split}
\vec v_B(t - \Delta t_B)= &
\vec v_B(t) 
- \frac{1}{4} \Delta t_B \frac{1}{m_B} \vec {f}_{B1}(\vec r_B, \vec v_B ; t ) \\
&- \frac{3}{4} \Delta t_B \frac{1}{m_B} 
\vec {f}_{B1}\bigg( \vec r_B -\frac{2}{3}\Delta t_B \vec v_B \\
&\qquad  , \vec v_B -\frac{2}{3} \frac{\Delta t_{B}}{m_B} \vec {f}_{B1}(t, \vec r_B, \vec v_B ) ; \\
&\qquad t - \frac{2}{3}\Delta t_B 
\bigg).
\end{split}
\end{equation}
In the evaluation of $\vec {f}_{B1}( , ;t - \frac{2}{3}\Delta t_B)$
it is required the knowledge of the value of the kinematical variables
of particle $A$ at retarded time $t - \Delta t'_A$ given by
\begin{equation}\label{eq:deltatap}
\begin{split}
&(t - \frac{2}{3}\Delta t_B  ) - (t -   \Delta t'_A) = \\
&\Delta t'_A - \frac{2}{3}\Delta t_B = 
|\vec r_B(t - \frac{2}{3}\Delta t_B) - \vec r_A(t- \Delta {t'}_A)|/c ;
\end{split}
\end{equation}
where we are taking
\begin{equation}\label{eq:rAendeltatp}
\vec r_A(t- \Delta {t'}_A)
=
\vec r_A(t) - \Delta {t'}_{A} \, \vec v_A(t) + \frac{\Delta {t'}_{A}^2}{2 m_A} \vec {f}_{A1}
.
\end{equation}
In the evaluation of $\Delta {t'}_{A}$ we can either use the iterative approach presented
previously, or we can solve exactly the quartic equation.

With this then we define the order two retarded force by
\begin{equation}\label{eq:fA-2}
\begin{split}
\vec {f}_{A2}(\vec r_A, \vec v_A ;t ) =& \vec {f}_A\bigg(\vec r_A , \vec v_A 
; \\
& \vec r_{B2}(t - \Delta t_{B1}), \\
& \vec v_{B2}(t - \Delta t_{B1}), \\
& \vec a_{B2}(t - \Delta t_{B1})
\bigg)
,
\end{split}
\end{equation}
where we estimate
\begin{equation}
\vec r_{B2}(t - \Delta t_{B1}) =
\vec r_B(t) - \Delta t_{B1} \, \vec v_B(t) + \frac{\Delta t_{B1}^2}{2 m_B} \vec {f}_{B1} ,
\end{equation}
\begin{equation}
\begin{split}
\vec v_{B2}(&t - \Delta t_{B1})= 
\vec v_B(t) 
- \frac{1}{4}  \frac{\Delta t_{B1}}{m_B} \vec {f}_{B1}(\vec r_B, \vec v_B ; t) \\
&- \frac{3}{4}  \frac{\Delta t_{B1}}{m_B} 
\vec {f}_{B}\bigg(
 \vec r_B -\frac{2}{3}\Delta t_{B1} \vec v_B , \\
&\qquad \qquad \vec v_B -\frac{2}{3}\frac{\Delta t_{B1}}{m_B} \vec {f}_{B1}(t, \vec r_B, \vec v_B), \\
&\qquad \qquad  \vec r_A(t - \Delta {t'}_{A}), \\
&\qquad \qquad  \vec v_A(t - \Delta {t'}_{A}), \\
&\qquad \qquad  \vec a_A(t - \Delta {t'}_{A})
\bigg),
\end{split}
\end{equation}
and
\begin{equation}
\vec a_{B2}(t - \Delta t_{B1})= \frac{1}{m_B} \vec {f}_{B1}(t, \vec r_B, \vec v_B ) ;
\end{equation}
with $r_A(t - \Delta {t'}_{A})$ given by (\ref{eq:rAendeltatp}),  and
\begin{equation}
\vec v_A(t - \Delta {t'}_{A}) = \vec v_A(t) - \frac{\Delta t'_{A}}{m_A} \, \vec {f}_{A1} ,
\end{equation}
and
\begin{equation}
\vec a_A(t - \Delta {t'}_{A}) = \frac{1}{m_A} \, \vec {f}_{A1} .
\end{equation}

Similarly, for particle $B$, one has the corresponding order two force given by
\begin{equation}\label{eq:fB-2}
\begin{split}
\vec {f}_{B2}(\vec r_B, \vec v_B ; t ) =& \vec {f}_B\bigg(\vec r_B , \vec v_B 
; \\
& \vec r_{A2}(t - \Delta t_{A1}) , \\
& \vec v_{A2}(t - \Delta t_{A1}) , \\
& \vec a_{A2}(t - \Delta t_{A1})
\bigg)
,
\end{split}
\end{equation}
where
\begin{equation}
\vec r_{A2}(t - \Delta t_{A1}) =
\vec r_A(t) - \Delta t_{A1} \, \vec v_A(t) + \frac{\Delta t_{A1}^2}{2 m_A} \vec {f}_{A1} ,
\end{equation}
\begin{equation}
\begin{split}
\vec v_{A2}(&t - \Delta t_{A1})= 
\vec v_A(t) 
- \frac{1}{4} \frac{\Delta t_{A1}}{m_A}  \vec {f}_{A1}(t, \vec r_A, \vec v_A ) \\
&- \frac{3}{4}  \frac{\Delta t_{A1}}{m_A} 
\vec {f}_{A}\bigg(
 \vec r_A -\frac{2}{3}\Delta t_{A1} \vec v_A \\
&\qquad \qquad \vec v_A -\frac{2}{3}\frac{\Delta t_{A1}}{m_A}  \vec {f}_{A1}(t, \vec r_A, \vec v_A ) , \\
&\qquad \qquad \vec r_B(t - \Delta {t'}_{B}), \\
&\qquad \qquad  \vec v_B(t - \Delta {t'}_{B}), \\
&\qquad \qquad  \vec a_B(t - \Delta {t'}_{B})
\bigg),
\end{split}
\end{equation}
and
\begin{equation}
\vec a_{A2}(t - \Delta t_{A1})= \frac{1}{m_A} \vec {f}_{A1}(\vec r_A, \vec v_A ;t ) ;
\end{equation}
with
\begin{equation}
\vec r_B(t - \Delta {t'}_{B}) = 
\vec r_B(t) - \Delta {t'}_{B} \, \vec v_B(t) + \frac{\Delta {t'}_{B}^2}{2 m_B} \vec {f}_{B1}
\end{equation}
\begin{equation}
\vec v_B(t - \Delta {t'}_{B}) = \vec v_B(t) - \frac{\Delta t'_{B}}{m_B} \, \vec {f}_{B1} ;
\end{equation}
and
\begin{equation}
\vec a_B(t - \Delta {t'}_{B}) = \frac{1}{m_B} \, \vec {f}_{B1} ;
\end{equation}
where $\Delta t'_{B}$ is the solution to
\begin{equation}\label{eq:deltatbp}
\begin{split}
&(t - \frac{2}{3}\Delta t_A  ) - (t -   \Delta t'_B) = \\
&\Delta t'_B - \frac{2}{3}\Delta t_A = 
|\vec r_A(t - \frac{2}{3}\Delta t_A) - \vec r_B(t- \Delta {t'}_B)|/c .
\end{split}
\end{equation}

We have just presented what constitutes the adaptation of the 
\emph{second order R-K method} to the case of a retarded system,
for the purpose of providing with an approximate single time set of
equations of motion with \emph{second order numerical accuracy}
for time steps of retarded times magnitude.
At each step, we have made full use of the previous calculated
quantities.
One can notice that the evaluation of the couple order two forces
requires the previous calculation of 
$\vec {f}_{A0}(\vec r_A, \vec v_A  ; t)$,
$\vec {f}_{B0}(\vec r_B, \vec v_B  ; t)$,
$\vec {f}_{A1}(\vec r_A, \vec v_A  ; t)$,
$\vec {f}_{B1}(\vec r_B, \vec v_B  ; t)$,
$\vec {f}_{A1}( , ;t - \frac{2}{3}\Delta t_A)$
and
$\vec {f}_{B1}( , ;t - \frac{2}{3}\Delta t_B)$;
and since the last two involve another couple of force
evaluation, they form a total
of eight evaluations of the force functions,
instead of the four evaluations one would have in the non-retarded case.
The extra evaluations are needed to maintain a 
\emph{second order numerical accuracy} of the retarded effects.

All this suggests the following remark.
If the retardations effects were small, then one would be tempted
to make a numerical evaluation of the dynamics in which the time
steps are of the same order of magnitude as the retarded times.
But the second order retarded forces respect the \emph{second order R-K}
scheme; which means that from the physical point of view it is
enough to carryout the numerical calculation with a 
\emph{second order numerical} scheme. 
Then, from this point of view, it is clear that the standard methods of
references \cite{Darwin1920}, \cite{Landau75} do not have 
a \emph{second order numerical} precision.
So, it seems that it would be a waste of computational resources to make a numerical integration
of equations of motions obtained from standard method\cite{Darwin1920,Landau75}
with a, let us say, \emph{fourth order integration scheme}; since
the dynamical equations where calculated with less numerical precision in the
retardation effects.
This is irrespective from the fact that given a first order ordinary
differential equation, the \emph{fourth order integration scheme}
will show better numerical properties than lower order ones, in general.

\section{Applying the model to the binary electromagnetic case}\label{sec:electro}

\subsection{The Lorentz force case}
The binary system of electromagnetic charges provides us with the opportunity to
study a couple of interesting physical systems.
The first one is provided by the system of interacting particle with
retarded fields through the Lorentz force.
This is the simplest relativistic binary system one can study, which
it can be applied to classical systems of particles with small charges.

In this case, each particle with charge $q$ generates the electromagnetic
field given by 
\begin{equation}
\begin{split}
F_{ab} &= 2 q \left( 
\frac{1}{r V} \hat l_{[a} \dot{\mathbf{v}}_{b]}
+
\frac{1}{r^2 V} ( 1 - \frac{r \dot V}{V}) \hat l_{[a}  \mathbf{v}_{b]}
\right) \\
&= 2 q \left( 
\frac{1}{r} \left[ l_{[a} \dot{\mathbf{v}}_{b]} - \frac{\dot V}{V} l_{[a}  \mathbf{v}_{b]}\right] 
+
\frac{1}{r^2}  l_{[a}  \mathbf{v}_{b]}
\right) 
;
\end{split}
\end{equation}
where we are using now a four dimensional notation, $a,b,...$ are abstract indices,
a dot means covariant derivative in the direction of the four velocity $\mathbf{v}^{b}$,
$l$ and $\hat l$ are proportional null vectors pointing from the retarded position
to the field point, $r$ is a retarded distance, $V=\hat l^\mu \mathbf{v}_\mu$
and we have chosen Gaussian units.
The notation is expanded in the Appendix.

Then, the forces for this system are
\begin{equation}
m_A \dot{\mathbf{v}}_A^a = q_A \, F(B)^a_{\;\;b} \, \mathbf{v}_A^b ;
\end{equation}
where $F^a_{\;\;b}(B)$ is the electromagnetic field tensor generated by particle $B$,
in which all quantities are evaluated at the corresponding retarded time.
Similarly one has, for particle $B$
\begin{equation}
m_B \dot{\mathbf{v}}_B^a = q_B \, F(A)^a_{\;\;b} \, \mathbf{v}_B^b .
\end{equation}
These are the four dimensional version of equations (\ref{eq:motA})
and (\ref{eq:motB}).
In the appendix we recall the notation to write these equations
of motion with a Galilean language.

We observe in this case that the force is linear in the single coupling
constant determined from the product of both charges.

This case is a paradigmatic example, since it has a couple of important
properties. 
Firstly, the field equations are linear in the intervening fields;
and it is clear then that the issue of precision in the calculation
of the retarded effects is completely unrelated to the subject
of the nature of the field equation; since for instance, in this case there is
no room for calculations of the field equations to higher orders.
Secondly, although the forces seem to involve only linearly the
coupling constant; the fact that involves the retarded accelerations
means that they are already hiding multiple apparitions of the
coupling constant, when the retarded effects are calculated
exactly.

\subsection{The back reaction force case}
If one wants to improve on the physical precision of the binary 
particle electromagnetic system, one must also consider the
effects of back reaction in the equations of motion due to 
the fact that accelerated charges emit electromagnetic
radiation.
In reference \cite{Gallo:2011tf} we have presented the most general
form of the equation of motion for charged particles which
balance the electromagnetic radiation generated by the motion.
We have argued there that the back reaction terms must be
understood in terms of orders in the evaluation of Lorentz
force. In particular we gave the explicit form of the forces
up to third order.
Let us recall here these forces.

We first define the four force vector
\begin{equation}
f_A^a = q_A \, F(B)^a_{\;\;b} \, \mathbf{v}_A^b ,
\end{equation}
and
\begin{equation}
{\bf f}_A^{2} \equiv - f_{A a} \; f_A^a .
\end{equation}
Then, we define the second order time derivative four vector $\dot f_{A(2)}$ from 
\begin{equation}
\dot f_{A(2)}^a = 
q_A \, \dot{ (F(B)^a_{\;\;b} ) } \, \mathbf{v}_A^b
+
q_A \, F(B)^a_{\;\;b} \, \frac{1}{m_A} f_A^b
.
\end{equation}

The third order equation of motion for particle $A$ is given by
\begin{equation}\label{eq:balance4m-3ord-b}
	m_A  \dot{\mathbf{v}}_A^a
	=
	f_A^b
	+ \frac{2}{3} q_A^2   (\ddot{\mathbf{v}}_{A(2)}^b - a_{A(2)}^2 \mathbf{v}_A^b  )
,
\end{equation}
where
\begin{equation}\label{eq:dotv2-b}
\dot{\mathbf{v}}_{A(2)}^b =  
\frac{1}{m_A } f_A^b
+   \frac{2}{3} q_A^2 
\frac{1}{m_A^2 }\dot f_{A(2)}^b
-
\frac{2}{3} q_A^2  
\frac{{\bf f}_A^2}{m_A^3}
\mathbf{v}_A^b
,
\end{equation}
and
\begin{equation}
a_{A(2)}^2 = 
-  \dot{\mathbf{v}}_{A(2)}^b \;  \dot{\mathbf{v}}_{A(2)b}
;
\end{equation}
and it is understood that in the evaluation of the right hand side of 
(\ref{eq:balance4m-3ord-b})
it is only required to maintain third order terns in ${\bf f}_A$.
The corresponding equation of motion for particle $B$ is obtained
from the above by interchanging the indices $A\longleftrightarrow B$
appropriately.

In this case we use a more sophisticated form of the equation of motion
with the objective to appropriately describe the effects of back
reaction in the motion due to the emission of electromagnetic radiation
of the accelerated charges.
These techniques involve several evaluations of the Lorentz force and
time derivatives of it; but it should be clear that these further evaluations
are needed to cover a specific effect, which is completely unrelated
to the need of evaluations of the forces to obtain a precise
single time dynamical model of a relativistic retarded system.

It has been customary in the literature to assume that the number of
evaluations of the forces should be universal for the construction
of a model; but we are here presenting arguments against
this assumption.

If we do not take into account the retardation effects with high enough
precision, in the construction of a
single time dynamical model of a relativistic retarded system,
then the model will not be able to accurately describe the dynamics
of the first order version of the force; and therefore
it would be meaningless to correct this dynamics by taking
higher order effects, as the back reaction effects discussed
in the case just presented.

\subsection{The single time dynamical model for binary charged particles}
We have just presented two theoretical models for a physical system consisting
of two charged particles; which in principle are exactly described in terms
of the two proper times of the particles. 
To each of these theoretical models we
can apply the techniques described previously, for the construction
of a single time dynamical model.
For a summary presentation of how to pass from the four dimensional
description to the Newtonian language, we refer to the Appendix.

It is our freedom to choose the degree of accuracy that we would
like to require to the single time model; and this choice normally depends
only on the nature of initial conditions, and/or on the evolution
of the system, but not on the nature
of the theoretical model one is using.

\section{Final comments}

We have presented above the order one retarded forces, and have shown that
they contain more dynamical information that the forces obtained by
the standard methods, as those of reference \cite{Darwin1920}.

We have also introduced the second order retarded forces; which has
even more dynamical information than the first order ones.

The discussions, in section \ref{sec:lagrangian}\,, on the 
possibility to construct a Lagrangian
from the equations of motion can be extended to higher order forces,
as those introduced in section \ref{sec:order2}\,; that we plan to carry out
elsewhere.

From the discussion presented above, concerning the degree of accuracy
of the single time equations of motion, it suggests that we
should revise the standard view that is applied to different
kind of approximations for the dynamics of compact objects;
since it is customary to base the studies in the choice of a universal
degree of approximation, normally based in a power of
interactive constants, and/or powers of the velocities.
This is normally suggested from the point of view of
the nature of field equations, as is done in the gravitational case;
and/or of from the nature of the equations of motion,
as is done in the electromagnetic case.
That is, it is customary to first choose a universal power
to be applied to any equation in the study, and then
determine the dynamics from it.
What we are suggesting with our previous discussion is that
one should choose the degree of accuracy one desires for
the evaluation of the retarded effects, independently from
anything else; that is, nature of the field equations
or any other independent physical consideration.
Since this choice for precision will determine the
degree in which the approximated forces will represent
the real global physical implications of the retardation
effects.

Our first example for the electromagnetic case, of
two particles interacting with retarded fields through
the Lorentz force, is a clear situation in which:
the exact field equations are linear,
and also the forces are linear in the fields.
However, we can ask for any desired precision in the calculation
of the retarded effects, what will be related with
several evaluations of the force function along the
trajectories, with the corresponding apparition 
of non-linear interactive terms in the final expressions for
the forces.
If one had used in this case the customary view point,
one would be forced to only use `zero order forces'(in our notation)
which will severely restrict the precision of the
final dynamical system.

The other electromagnetic case considered in section \ref{sec:electro} reinforces
the view that the degree needed for the appropriate calculation
of different effects must be considered separately.
This of course is in contrast to the customary attitude which
considers a universal choice of order first, and then
proceed with the calculation.

The consideration of binary gravitating systems also involve the issue of
the retarded effects; however the retarded effects are rarely considered
separately.
For example in the seminal article \cite{Einstein:1938yz} that derive
the first post-Newtonian equations; they have considered instantaneous
accelerations.
In classical approaches to the post-Newtonian framework each potential is calculated
from Poisson equation, so that each potential is necessarily
instantaneous, and the retardation effects are implicit and hidden from 
view\cite{Poisson2014}.
And to this one must add that since the usual philosophy is to choose
a post-Newtonian order (PN), and then calculate all the dynamics
in terms of this choice, these frameworks do not incorporate
the retarded effects we have presented here.
Naturally, when considering gravitating systems, the problem
of the retarded effects complicates considerably when 
higher order geometries are taken into account; since
the past null cones are calculated in the corresponding
curved spacetime.

Although we have presented the order one and order two set of equations;
it is clear that the procedure presented here can be extended to any
order of accuracy one desires. This can be done, for example, by taking a standard
nth order Runge-Kutta method, and at each stage use the previous elements
to calculate the needed retarded times; in analogy to what we
have presented here.
As we have seen, to obtain a nth R-K order of accuracy we need
about 2*n number of force evaluations for each particle.

It is worthwhile to remark, that
our work gives an answer to the unsolved problem stated in \cite{Havas:1962zz};
namely we give a constructive way for the initial value problem
of a set of relativistic particles with mechanical initial data,
i.e. position and velocities, with arbitrary desired precision.

We plan to apply this discussion to numerical calculation of binary
charged particle systems and to
the problem of equations of motion
for gravitating systems in future works.

\subsection*{Acknowledgments}

We thank Emanuel Gallo for a careful reading of the manuscript,
for comments and suggestions.

We acknowledge support from CONICET, SeCyT-UNC and Foncyt.

\appendix
\section{Relations among coordinates and null 
vectors}

\subsection{The inertial system}

Let us denote with  $y^\mu$  the standard Cartesian coordinates and with  
$(\hat x^0 = \hat u,\hat x^1 = \hat r,\hat x^2,\hat x^3)$,
where $(\hat x^2,\hat x^3)=(\theta,\phi)$ or $\zeta =\frac{\hat x^2 + i \hat x^3}{2}$,
the corresponding null polar coordinates; then, the relation 
between them is given by
\begin{equation}\label{eq:nulpolar}
y^\mu = \hat u \, \delta^\mu_0 + \hat r \, \hat l^\mu(\zeta, \bar \zeta )
;
\end{equation}
with
\begin{equation}
 \hat l^\mu(\zeta, \bar \zeta ) = \hat l_0^\mu(\zeta, \bar \zeta ) ;
\end{equation}
defined 
by
\begin{equation}  \label{eq:bflhat}
\begin{split}
\hat l_0^\nu(x^2,x^3)
\equiv &
\biggl(1,\sin(\theta) \cos(\phi), \sin(\theta) \sin(\phi), \cos(\theta) \biggr)\\
=&\left( 1,\frac{\zeta +\bar \zeta }{
1+\zeta \bar{\zeta }},\frac{\zeta -\bar{\zeta }}{i(1+\zeta
\bar{\zeta )}},\frac{\zeta \bar{\zeta }-1}{1+\zeta \bar{\zeta
}}\right) \\
=&
\bigg(
\sqrt{4\pi} Y_{0,0} , 
- \sqrt{\frac{2\pi}{3}} \big(Y_{1,1} - Y_{1,-1} \big) , \\
& i \sqrt{\frac{2\pi}{3}} \big(Y_{1,1} + Y_{1,-1} \big) ,
\sqrt{\frac{4\pi}{3}} Y_{1,0}
\bigg)
;
\end{split}
\end{equation}
where $\mu, \nu, \cdots =0,1,2,3$,
and we are using either the standard sphere angular coordinates 
$(\theta,\phi)$ or the complex angular coordinates $\zeta =\frac{\hat x^2 + i \hat x^3}{2}$,
where $(\zeta,\bar\zeta)$ are complex stereographic coordinates
of the sphere; which are related to the standard coordinates by
$\zeta = e^{i \phi} \cot(\frac{\theta}{2})$.
The $Y_{l,m}$ are the usual spherical harmonic on the sphere.

\subsection{The intrinsic non-inertial system}

Let $z^\mu(\tau_0)$ be the evolution of the particle,
in terms of a Cartesian coordinate system,
 with proper time $\tau_0$.
We define a null function $u'$ as the future null cones emanating from
$z^\mu(\tau_0)$, such that $u'= \tau_0$ at the world line of the particle.

If  $(x^0\! =\! u',x^1\! =\! r, x^2, x^3)$,
where $(x^2,x^3)=(\theta',\phi')$ or $\zeta' =\frac{x^2 + i x^3}{2}$,
are the null polar coordinates adapted to an arbitrary 
timelike curve, determined by  $z(u')^\mu$, then one has
\begin{equation}\label{eq:nulpolar'}
y^\mu= z^\mu(u')+r \, \mathbf{l}^\mu(u,\zeta', \bar \zeta' ) 
.
\end{equation}

Note that
\begin{equation}
(y^\mu  - z^\mu(u')) \mathbf{l}_\mu  = r \, \mathbf{l}^\mu \mathbf{l}_\mu = 0 ,
\end{equation}
and that
\begin{equation}
(y^\mu  - z^\mu(u')) \mathbf{v}_\mu  = r \, \mathbf{l}^\mu \mathbf{v}_\mu = r ;
\end{equation}
so that
\begin{equation}\label{eq:boldl}
 \mathbf{l}^\mu = \frac{y^\mu  - z^\mu(u')}{(y^\nu  - z^\nu(u')) \mathbf{v}_\nu} .
\end{equation}

Given a fixed point $y^\mu$ one has to take different spacelike directions,
and therefore different angular coordinates, for the two null vectors
to reach the fixed point.
But, if given a particular future null cone determined by the apex $z(u')$,
one also chooses an inertial frame with origin at this apex, then, from
equations (\ref{eq:nulpolar}) and (\ref{eq:nulpolar'}) one deduces that
at this cone the two null vectors $\mathbf{l}^\mu(u,\zeta', \bar\zeta')$
and $\hat l_0^\mu(\zeta', \bar\zeta')$ must be proportional;
so that
\begin{equation}
 \mathbf{l}^\mu(u,\zeta', \bar \zeta' ) 
= \alpha(u,\zeta', \bar \zeta' ) \hat l_0^\mu(\zeta', \bar \zeta' ) ;
\end{equation}
with $\alpha>0$.
But then
we have
\begin{equation}
 1 = \mathbf{l}^\mu \mathbf{v}_\mu 
= \alpha \hat l_0^\mu \mathbf{v}_\mu
;
\end{equation}
which implies that
\begin{equation}\label{eq:v-eta-1}
\frac{1}{\alpha}  
=  V
,
\end{equation}
with
\begin{equation}\label{eq:v-eta}
V \equiv  \hat l_0^\mu \mathbf{v}_\mu
,
\end{equation}
and also
\begin{equation}\label{eq:lconlhat}
\mathbf{l}^\mu(u,\zeta',\bar\zeta') 
= \frac{1}{V(u,\zeta',\bar\zeta')} \hat l_0^\mu(\zeta',\bar\zeta')
.
\end{equation}

\subsection{Basic relations for coordinate velocities and accelerations}

We use for the four velocity the notation
\begin{equation}
\mathbf{v}^\mu = \frac{d x^\mu}{d\tau_0} ;
\end{equation}
where $\tau_0$ is the proper time with respect to the flat metric,
and
\begin{equation}
\vec v = \big( v^i\big) = \big(\frac{d x^i}{dt}\big),
\end{equation}
for $i=1,2,3$ and $t=x^0$.
Then, one has
\begin{equation}
\mathbf{v}^\mu = \frac{1}{\sqrt{1 - v^2}} (1,v^i)
= \Gamma (1,\vec v)
;
\end{equation}
where we use the notation $\Gamma = \frac{dt}{d\tau_0} = \frac{1}{\sqrt{1 - v^2}}$
and $v^2 = \vec v \cdot \vec v$.
With this notation one can also express the four acceleration as
\begin{equation}
\mathbf{a}^\mu = \frac{d \mathbf{v}^\mu}{d\tau_0}
= \Gamma \frac{d \mathbf{v}^\mu}{dt}
= \Gamma \frac{d\Gamma}{dt} (1,\vec v)
+ \Gamma^2 (0,\vec a)
;
\end{equation}
where
\begin{equation}
\vec a = \big( a^i\big) =  \big( \frac{dv^i}{dt}\big) = \frac{d \vec v}{dt}
.
\end{equation}
Note that
\begin{equation}
\frac{d\Gamma}{dt} = \Gamma^3 v \frac{dv}{dt}
,
\end{equation}
and
\begin{equation}
v \frac{dv}{dt} = \frac{1}{2} \frac{d}{dt}\vec v \cdot \vec v
= \vec a \cdot \vec v
;
\end{equation}
so that one has
\begin{equation}
\begin{split}
\mathbf{a}^\mu 
=& \Gamma^4 v \frac{dv}{dt} (1,\vec v) + \Gamma^2 (0,\vec a) \\
=& \Gamma^4 (\vec a \cdot \vec v) (1,\vec v) + \Gamma^2 (0,\vec a) \\
=& \bigg(\Gamma^4 (\vec a \cdot \vec v), \Gamma^4 (\vec a \cdot \vec v) \vec v + \Gamma^2 \vec a  \bigg)  
;
\end{split}\end{equation}

We also use the notation
\begin{equation}
\mathbf{a}^\mu \mathbf{a}_\mu =  - \mathbf{a}^2
;
\end{equation}
so that $\mathbf{a}^2$ is a positive quantity.

From the  point of view of equations of motion, the physical important quantity
is the momentum of the particle defined by
\begin{equation}
p^\mu = m \mathbf{v}^\mu ,
\end{equation}
and the equation of motion is written in the form
\begin{equation}
\frac{d p^\mu}{d\tau} = f^\mu .
\end{equation}
In the case of the Lorentz force one can write
\begin{equation}
\frac{d p^\mu}{d\tau} = \Gamma \frac{d p^\mu}{dt}
= q  \, F^\mu_{\;\;\nu} \, \mathbf{v}^\nu ;
\end{equation}
therefore for the spacelike components we can write
the equation of motion in the form
\begin{equation}
\frac{d p^i}{dt}
= q  \, F^i_{\;\;\nu} \, \frac{1}{\Gamma} \mathbf{v}^\nu ;
\end{equation}
where we have seen that 
$\frac{1}{\Gamma} \mathbf{v}^\nu = (1,\vec v)$;
so that
\begin{equation}
\frac{d p^i}{dt}
= q  \, \bigg( E^i + ( \vec v \times \vec B )^i \bigg);
\end{equation}
which is the standard way to write the Lorentz force in terms
of the three dimensional variable.
And using that $p^i = \Gamma m v^i$, one can also write
\begin{equation}
\frac{d \vec p}{dt} = \Gamma m \vec a + \frac{d\Gamma}{dt} m \vec v
= \Gamma m \vec a + \Gamma^3 (\vec a \cdot \vec v) m \vec v
;
\end{equation}
so that in terms of the standard acceleration Newtonian form
one can express the equation of motion as
\begin{equation}
m \vec a =
 \frac{q}{\Gamma} \, \bigg(\vec E +  \vec v \times \vec B  \bigg)
- \Gamma^2 (\vec a \cdot \vec v) m \vec v
;
\end{equation}
from which we can see that
\begin{equation}
m \, \vec a \cdot \vec v \bigg(1 + \Gamma^2   v^2 \bigg) =
m \, \vec a \cdot \vec v \; \Gamma^2  =
\frac{q}{\Gamma} \, \vec E \cdot \vec v 
;
\end{equation}
so that the final equation of motion in Newtonian notation is
\begin{equation}
m \vec a =
\frac{q}{\Gamma} \, \bigg(\vec E +  \vec v \times \vec B  \bigg)
- \frac{q}{\Gamma} \, \big( \vec E \cdot \vec v \big)
\vec v
;
\end{equation}
which is seldom shown explicitly in textbooks\cite{Landau71}.

%
%
%

\end{document}